\documentclass[prb,reprint,amsmath,amssymb,aps,superscriptaddress]{revtex4-2}

\usepackage{amsmath,amsfonts,amssymb} %
\usepackage{graphicx}
\graphicspath{{figs/}}
\usepackage{dcolumn}
\usepackage{bm}
\usepackage[acronym]{glossaries}
\usepackage{tabularx} %
\usepackage{braket} %
\usepackage{multirow}
\usepackage[dvipsnames,table]{xcolor} %
\usepackage[breaklinks, %
colorlinks, linkcolor=Blue, citecolor=Blue, urlcolor=Blue]{hyperref} %
\usepackage{booktabs} %

\newcommand{\eq}[1]{Eq.~(\ref{#1})} %
\newcommand{\eqs}[1]{Eqs.~(\ref{#1})} %
\def\be{\begin{equation}} %
\def\ee{\end{equation}} %
\newcommand{\bea}{\begin{eqnarray}}
\newcommand{\eea}{\end{eqnarray}}
\newcommand{\sparen}[1]{\bigl(#1\bigr)} %
\newcommand{\var}[1]{\text{Var}\left(#1\right)}
\newcommand{\varq}[2]{\text{Var}_{#2}\left(#1\right)}
\newcommand{\cov}[1]{\text{Cov}\left(#1\right)}
\newcommand{\covq}[2]{\text{Cov}_{#2}\left(#1\right)}
\newcommand{\fracpar}[2]{\frac{\partial #1}{\partial #2}} %

\begin{document}
    \setcitestyle{super} 
    \title{Deterministic improvements of quantum measurements with grouping of compatible operators, non-local transformations, and covariance estimates}

    \author{Tzu-Ching Yen}
    \affiliation{Chemical Physics Theory Group, Department of Chemistry, University of Toronto, Toronto, Ontario M5S 3H6, Canada}

    \author{Aadithya Ganeshram}
    \affiliation{Chemical Physics Theory Group, Department of Chemistry, University of Toronto, Toronto, Ontario M5S 3H6, Canada}

    \author{Artur F. Izmaylov}
    \affiliation{Chemical Physics Theory Group, Department of Chemistry, University of Toronto, Toronto, Ontario M5S 3H6, Canada}
    \affiliation{Department of Physical and Environmental Sciences, University of Toronto Scarborough, Toronto, Ontario M1C 1A4, Canada}

    \date{\today}

    \begin{abstract}
    Obtaining the expectation value of an observable on a quantum computer is a crucial step in the variational quantum algorithms. For complicated observables such as molecular electronic Hamiltonians, a common strategy is to present the observable as a linear combination of measurable fragments. The main problem of this approach is a large number of measurements required for accurate estimation of the observable's expectation value. We consider several partitioning schemes based on grouping of commuting multi-qubit Pauli products with the goal of minimizing 
   the number of measurements. Three main directions are explored: 1) grouping commuting operators using the greedy 
   approach, 2) involving non-local unitary transformations for measuring, 
   and 3) taking advantage of compatibility of some Pauli products with several 
   measurable groups. The last direction gives rise to a general framework that not only provides improvements over previous methods but also connects measurement grouping approaches with recent advances in techniques of shadow tomography. 
    Following this direction, we develop two new measurement schemes that achieve a severalfold reduction in the number of measurements for a set of model molecules compared to previous state-of-the-art methods. 

    \end{abstract}

    \maketitle
        \section{Introduction}
        Variational Quantum Algorithms (VQA) constitute one of the most promising class of applications for quantum computers in the noisy intermediate scale quantum era. \cite{Preskill:2018jf,Peruzzo2014}
    In VQAs, classically intractable optimization problems are represented as lowest eigenstates of  $N_q$-qubit operators 
    \bea
        \hat H = \sum_{n=1}^{N_P} c_n \hat P_n, ~\hat P_n = \otimes_{k=1}^{N_q} \hat \sigma_k
    \eea
    where $c_n$'s are coefficients and $\hat P_n$'s are tensor products of Pauli operators or identities, $\hat \sigma_k \in \{\hat x_k, \hat y_k, \hat z_k, \hat 1_k\}$.
    VQAs then solve these problems by minimizing $
        E(\boldsymbol{\theta}) = \bra{\psi\left(\boldsymbol{\theta}\right)} \hat H \ket{\psi\left(\boldsymbol{\theta}\right)},
    $ where the quantum computer prepares the trial wavefunction $\ket{\psi\left(\boldsymbol{\theta}\right)}$ and is given a task to measure $E(\boldsymbol{\theta})$, while a classical optimizer determines the optimal $\boldsymbol{\theta}$.
However, it was found that estimating $E(\boldsymbol{\theta})$ accurately for chemical systems requires large numbers of measurements that diminish VQA's advantage over classical alternatives.\cite{gonthier2020identifying} 

    Measuring $E(\boldsymbol{\theta})$ is indeed not a straightforward task since only $z$-Pauli operators can be measured on current digital quantum computers. 
    A common approach to measuring the expectation value of the Hamiltonian is to present $\hat H$ as a sum of measurable fragments $\hat H = \sum_\alpha \hat A_\alpha$. The condition for selecting $\hat A_\alpha$'s is that they can be easily rotated into polynomial functions of $z$-Pauli operators 
    \bea\label{eq:mfrag}
        \hat A_\alpha = \hat U_\alpha^\dagger \left[ \sum_i a_{i,_\alpha} \hat z_i + \sum_{ij} b_{ij,\alpha} \hat z_i \hat z_j + ...\right] \hat U_\alpha.
        \label{eq:measurable_frag}
    \eea 
    Then $\bra{\psi\left(\boldsymbol{\theta}\right)} \hat H \ket{\psi\left(\boldsymbol{\theta}\right)} = \sum_\alpha \bra{\psi\left(\boldsymbol{\theta}\right)} \hat A_\alpha \ket{\psi\left(\boldsymbol{\theta}\right)}$
    where the latter can be obtained by measuring $z$-Pauli operators of $\hat A_\alpha$ for the rotated wavefunction $\hat U_\alpha\ket{\psi\left(\boldsymbol{\theta}\right)}$.

    Unfortunately, in general, the wavefunction $\ket{\psi\left(\boldsymbol{\theta}\right)}$ is not an eigenstate of $\hat A_\alpha$, and thus each fragment requires 
    a set of measurements 
 to obtain an estimator $\bar{A}_\alpha$ for $\bra{\psi\left(\boldsymbol{\theta}\right)} \hat A_\alpha \ket{\psi\left(\boldsymbol{\theta}\right)}$.
    The efficiency of the Hamiltonian measurement scheme is determined by the total number 
    of measurements, $M$, 
    needed to reach $\epsilon$ accuracy for $E(\boldsymbol{\theta})$. For a
    simple estimator of $E(\boldsymbol{\theta})$ as the sum of $\bar{A}_\alpha$ estimators,
    the error scales as $\epsilon = \sqrt{\sum_{\alpha} \varq{\hat A_\alpha}{\psi}/m_\alpha}$, where $\varq{\hat A_\alpha}{\psi} = \bra{\psi} \hat A_\alpha^2 \ket{\psi} - \bra{\psi} \hat A_\alpha \ket{\psi}^2$ is the variance of 
    each fragment, and $m_\alpha$'s are the numbers of measurements allocated for each fragment, with the condition $\sum_\alpha m_\alpha = M$. The optimal distribution of 
    measurements is $m_\alpha \sim \sqrt{\varq{\hat A_\alpha}{\psi}}$,
    which gives the total estimator error as $\epsilon = \sum_{\alpha} \sqrt{\varq{\hat A_\alpha}{\psi}}/\sqrt{M}$.

    This consideration shows superiority of estimators operating with a set of measurable fragments that have the lowest sum over variance square roots. For practical use of this consideration, there are two difficulties in explicit optimization of 
    the estimator error: 1) there is an 
    overwhelming number of choices for measurable operator fragments and 2) variance estimates require knowledge of the wavefunction. The second problem can be addressed by introducing a classically efficient proxy for the quantum wavefunction (e.g. from Hartree-Fock or configuration interaction singles and doubles (CISD) methods in quantum chemistry problems) or by utilizing the measurement results from VQAs to gain empirical estimates. 
    Yet, the search space in the first problem is so vast that it has only been addressed heuristically in previous studies. 
    The Hamiltonian partitioning has been done in qubit space\cite{Izmaylov2019revising,Verteletskyi:2020do,MoscaA,Yen2019b,Gokhale_ON3,Izmaylov2019unitary,Zhao2019PRA, Hamamura_2020} and in the original fermionic space with subsequent transfer of all operators into the qubit space.\cite{huggins2019efficient,Yen:2021PRX} An initial heuristic idea was to reduce the number of measurable fragments without accounting for variances. It was shown for several partitioning that the number of fragments is not a good proxy for the total number of measurements, and the fragments' variances cannot be ignored.\cite{Crawford2021efficientquantum,Yen:2021PRX} 
    The key element determining a particular set of measurable fragments is a class of unitary transformations $\hat U_\alpha$'s in \eq{eq:mfrag}. Compared to single-qubit transformations, multi-qubit transformations are more flexible and therefore
    have a greater potential to minimize the total number of measurements by 
    selecting fragments with lower variances. Yet, they also have a downside of an 
    extra circuit overhead needed to perform the rotation before the measurement. 
    Once the set of unitary transformations has been selected, empirically, it was found 
    more beneficial for the estimator variance to use greedy algorithms for the Hamiltonian partitioning. In these algorithms one finds $\hat A_\alpha$ fragments sequentially 
    by minimizing the norm of the difference between partial sum of $\hat A_\alpha$'s and $\hat H$.\cite{Crawford2021efficientquantum,Yen:2021PRX} 
    This can be rationalized considering that greedy algorithms produce first 
    fragments with larger variances and later fragments with smaller variances. Such a distribution of variances makes sum of square roots somewhat smaller compare to the case where variances are distributed relatively equally over all fragments. 

    Fragmentation techniques in the qubit space are based on grouping mutually commuting Pauli products 
    in each fragment $\hat A_\alpha$ [\eq{eq:mfrag}]. Two types of commutativity between Pauli products 
    are used: qubit-wise and full commutativity.
    The full commutativity (FC) is the 
    regular commutativity of two operators,\cite{Yen2019b} 
    whereas the qubit-wise commutativity (QWC) for two Pauli products is a condition when 
    corresponding single-qubit operators commute.\cite{Verteletskyi:2020do}
    Using either commutativity to find $\hat A_\alpha$'s, one can efficiently identify unitary 
    operators $\hat U_\alpha$'s from the Clifford group 
    that bring the fragments to the form of \eq{eq:measurable_frag} for measurement.
    Only one-qubit Clifford gates are sufficient for $\hat U_\alpha$'s of the qubit-wise commuting fragments,\cite{Verteletskyi:2020do} while $\hat U_\alpha$'s for fully commuting 
    fragments require also two-qubit Clifford gates.\cite{Yen2019b} 
    
\begin{figure}[h!]
  \includegraphics[width=1\columnwidth]{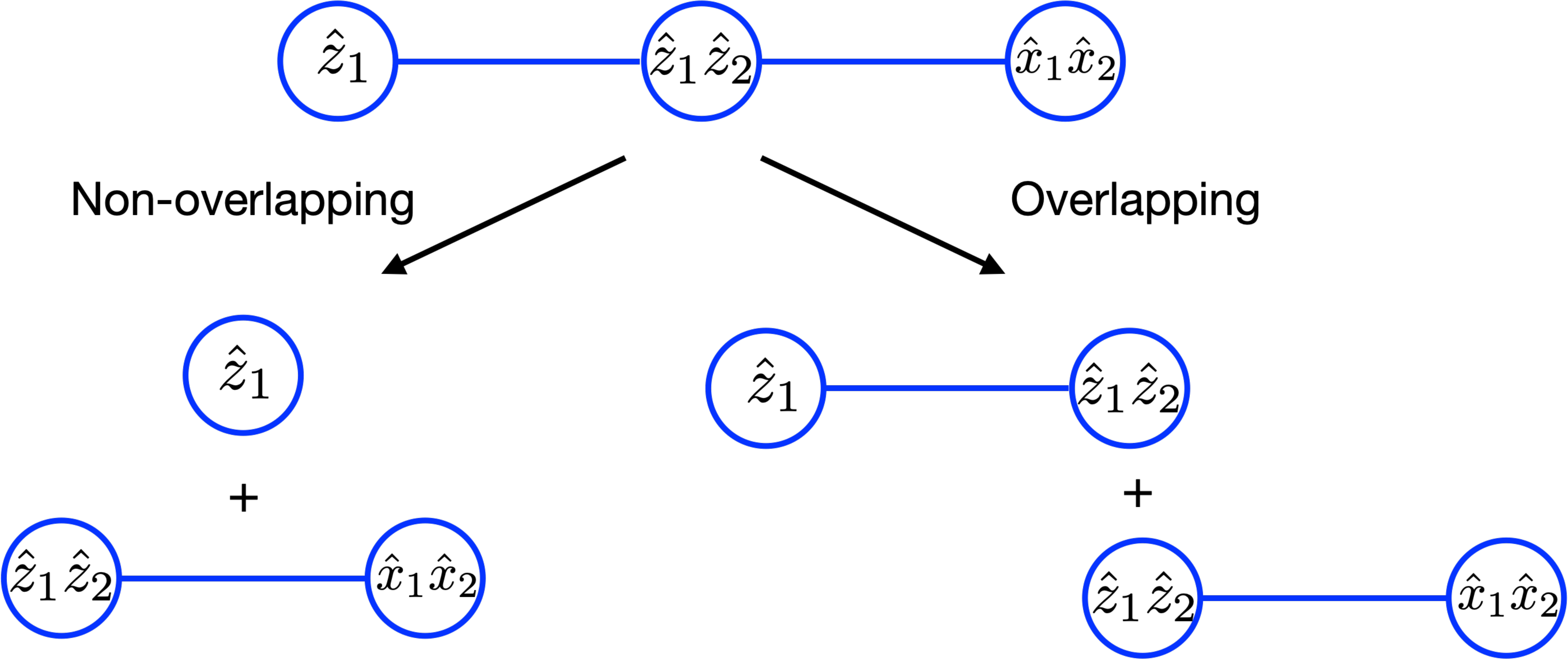}
  \caption{Illustration of non-overlapping and overlapping partitioning based on full commutativity for a model Hamiltonian, $\hat H = c_1 \hat z_1 + c_2 \hat z_1 \hat z_2 + c_3 \hat x_1 \hat x_2$. Within the non-overlapping scheme the fragments are: $\hat A_1 = c_1 \hat z_1$ and $\hat A_2 = c_2 \hat z_1 \hat z_2 + c_3 \hat x_1 \hat x_2$. For the overlapping scheme of Sec.~\ref{sec:CS} (Sec.~\ref{sec:MA}) the fragments are: $\hat A_1 = c_1 \hat z_1+c_2^{(1)} \hat z_1 \hat z_2$ ($\hat A_1 = c_1 \hat z_1+c_2 \hat z_1 \hat z_2$) and $\hat A_2 = c_2^{(2)} \hat z_1 \hat z_2 + c_3 \hat x_1 \hat x_2$ ($\hat A_2 = c_2 \hat z_1 \hat z_2 + c_3 \hat x_1 \hat x_2$).
  }
  \label{fig:NoO}
\end{figure}

Initial QWC- and FC-based schemes had $\hat A_\alpha$'s consisting of disjoint (non-overlapping) 
sets of Pauli products. Generally, each Pauli product can belong to multiple $\hat A_\alpha$'s as long
as it commutes with all terms in these fragments. This follows from non-transitivity of both FC and 
QWC as binary relations: if $\hat P_1$ commutes with $\hat P_2$, and $\hat P_2$ commutes with $\hat P_3$, this does not lead to commutativity of $\hat P_1$ and $\hat P_3$. 
For the measurement problem, $\hat P_1$ and $\hat P_3$ form separate measurable groups 
while $\hat P_2$ can be measured within both of these groups. Here, $\hat P_2$ constitutes 
an overlapping element for the $\hat P_1$ and $\hat P_3$ groups (see Fig.~\ref{fig:NoO} where $\hat P_1$, $\hat P_2$, and $\hat P_3$ are  $\hat z_1$, $\hat z_1 \hat z_2$, and  $\hat x_1 \hat x_2$ respectively).
Recent developments based on shadow tomography 
\cite{hadfield2020_locallyBiased, Huang_2021_derand,hillmich2021decision,hadfield2021adaptive} 
and grouping\cite{wu2021overlapped,shlosberg2021adaptive} techniques 
exploiting overlapping fragments found considerable reduction in the number of needed measurements over non-overlapping grouping schemes. However, all non-overlapping schemes used in those comparisons 
did not use the greedy approach. 
Since within qubit-based partitioning schemes there are multiple estimator improvement techniques, it is interesting to assess them all systematically.  

In this work, we assess improvements in the total number of measurements from introducing 
a series of ideas: 1) grouping commuting operators using the greedy approach, 2) involving non-local (entangling) unitaries for measuring groups of fully commuting Pauli products, and 3) taking advantage of compatibility of some Pauli products with several measurable groups (i.e. overlapping grouping). It is shown that these ideas, used separately or combined, can give rise to
schemes superior to prior art within grouping and shadow tomography techniques. One of the most striking findings is that using only greedy non-overlapping grouping within the QWC approach can already surpass the performance of recent shadow tomography techniques that employed overlapping local frames.
We do not consider fermionic-algebra-based techniques here because they do not allow overlapping grouping 
while all other improvements were already discussed for them.\cite{Yen:2021PRX} 
Other measurement techniques that do not involve grouping of Hamiltonian terms
are also outside of the scope of the current work. \cite{radin2021classicallyboosted_arxiv, Kyriienko_PRXQ_HApprox, Wang_PRXQ_MinRuntime, Torlai_PRR_NeuralNetworkMeas,PRXQuantum:POVM}

    \section{Theory}
    
    
    \subsection{Estimator for non-overlapping Pauli groups}

    All measurable fragments $\hat A_\alpha$'s are linear combinations of mutually commuting or qubit-wise commuting Pauli products
    \bea
        \hat A_\alpha &=& \sum_{k} c_k \hat P_k,\ \hat P_k \in \mathcal P_\alpha,
    \eea
    where $\mathcal P_\alpha$'s are disjoint sets of Pauli products measured as parts 
    of corresponding $\hat A_\alpha$'s, and $c_k$'s are coefficients of 
    $\hat P_k$'s in the Hamiltonian.    
    The commutativity between Pauli products within $\mathcal P_\alpha$ implies a common 
    eigen-basis $\mathbf{B}_\alpha$, where one can measure all the members of $\mathcal P_\alpha$.
    Initial proposals to find these fragments aim to minimize the total number of fragments using graph coloring algorithms, such as the largest first (LF) algorithm. \cite{Verteletskyi:2020do, Yen2019b} 
    But later the sorted insertion (SI) algorithm employing the greedy approach was found to produce lower variances for the energy estimator. \cite{Crawford2021efficientquantum} 

    Let $\bar{H}$ denote the estimator for $\bra{\psi} \hat H \ket{\psi}$; it is a sum of estimators for its parts 
    \bea
        \bar{H} &=& \sum_{\alpha=1}^L \bar{A}_\alpha. 
    \eea
    Each $\bar A_\alpha$ comes from $m_\alpha$ repeated measurements of $\hat A_\alpha$,
    \bea
    \bar{A}_\alpha &=& \frac{1}{m_\alpha} \sum_{i=1}^{m_\alpha} 
        A_{\alpha, i },
    \eea where $A_{\alpha, i}$ is the $i$-th measurement result of $\hat A_\alpha$. 
    The variance of $\bar{H}$ is 
    \bea
        \var{\bar H} &=& \sum_{\alpha=1}^L \var{\bar A_\alpha},
    \eea
    where $\var{\bar A_\alpha}$'s are variances of estimators characterizing differences between $\bar{A}_\alpha$'s and the true expectation values $\bra{\psi} \hat A_\alpha \ket{\psi}$'s. Note that covariances between different fragments ${\rm Cov}(\bar{A}_\alpha,\bar{A}_\beta)$ are zero because measurements of different 
    fragments are done independently. 
    $\var{\bar A_\alpha}$'s can be evaluated using quantum operator 
    variances $\varq{\hat A_\alpha}{\psi}$'s, $\var{\bar A_\alpha} = \varq{\hat A_\alpha}{\psi} /m_\alpha$, which leads to the Hamiltonian estimator variance as 
    \bea
            \var{\bar H} = \sum_{\alpha=1}^L \frac{1}{m_\alpha} \varq{\hat A_\alpha}{\psi}.
        \label{nonoverlapVar}
    \eea
    Using the constraint $M = \sum_\alpha m_\alpha$ one can minimize $\var{\bar H}$ with respect to $m_\alpha$'s\cite{Rubin_2018, Crawford2021efficientquantum} 
    \bea
        m_\alpha &=& \sqrt{\varq{\hat{A}_\alpha}{\psi}} \frac{\sum_\beta \sqrt{\varq{\hat{A}_\beta}{\psi}}}{\var{\bar H}} 
        \label{eq:nonoverlapDistribution} \\
        \var{\bar H} &=& \frac{1}{M} \left( \sum_\alpha \sqrt{\varq{\hat{A}_\alpha}{\psi}}\right)^2.\label{eq:Hnonov}
    \eea 
    
    In practice, quantum variances $\varq{\hat A_\alpha}{\psi}$ are not known {\it a priori}. They can be evaluated using covariances between Pauli products,
    \bea
    \hspace{-2cm}
        \varq{\hat{A}_\alpha}{\psi} &=& \sum_{jk} c_j c_{k} \covq{\hat P_j,  \hat P_{k}}{\psi} \\ \notag
        \covq{\hat P_j, \hat P_{k}}{\psi} &=& \bra{\psi} \hat P_j \hat P_{k} \ket{\psi} -
        \bra{\psi} \hat P_j \ket{\psi} \\
        &&\times \bra{\psi} \hat P_{k} \ket{\psi}, 
    \eea where $\hat P_j,\ \hat P_{k} \in \mathcal P_\alpha$. The covariances 
    for different Pauli products are generally non-zero because all of these Pauli products are measured together within the same fragment. The covariances can be approximated for molecular Hamiltonians using approximate wavefunctions obtained on a classical computer. 
   Configuration interaction singles and doubles (CISD) is one example for obtaining approximation for $\ket{\psi}$ that will be used in the current work. 
   Alternatively, the measurements results obtained from measurement basis $\mathbf{B}_{\alpha}$ can help estimate the covariances between Pauli products of $\mathcal P_\alpha$ during VQA's cycles. 

    \subsection{Optimization by Coefficient Splitting}
    \label{sec:CS}
    
    Many Pauli products in the Hamiltonian can be measured in multiple fragments because of their compatibility with other members of those fragments. 
    The coefficient splitting approach, briefly mentioned in Ref.~\citenum{Crawford2021efficientquantum}, 
    takes advantage of this opportunity by splitting coefficients of 
    Pauli products that are compatible with multiple fragments 
    \bea
        \hat A_\alpha &=& \sum_{k} c_k^{(\alpha)} \hat P_k,\ \hat P_k \in \mathcal P_\alpha\\
        c_k &=& \sum_{\alpha \in \mathcal I_k} c^{(\alpha)}_k 
        \label{eq:cs_constraint}
    \eea where $\mathcal{I}_k$ is a set of group indices $\alpha$ corresponding to fragments $\hat {A}_\alpha$ whose members are compatible with $\hat P_k$ (see Fig.~\ref{fig:NoO} for an example). 
    Note that the equations for the estimator variance and the optimal measurement distribution remain the same [\eqs{eq:nonoverlapDistribution} and \eqref{eq:Hnonov}].
    However, freedom in the coefficient splitting approach [\eq{eq:cs_constraint}] can be used to 
    minimize the Hamiltonian estimator variance [\eqref{eq:Hnonov}].
    
    
    A straightforward approach to minimization of $\var{\bar H}$ with respect to 
     $c^{(\alpha)}_k$'s is to use analytical gradients $\partial \var{\bar H}/\partial c^{(\alpha)}_k$. The gradients are non-linear functions of $c^{(\alpha)}_k$ and computing them
     becomes computationally expensive 
     as the number of $c^{(\alpha)}_k$'s grows with the size of the system. As a computationally more efficient alternative, we propose an iterative heuristic that quickly converges to a zero gradient solution. 
    %
    
    {\it Iterative coefficient splitting (ICS): }
    Given a particular choice of $ c^{(\alpha)}_k $'s and its optimal $m_\alpha$'s, 
    the procedure consists of iteratively applying two steps: 
    (1) optimization of $ c^{(\alpha)}_k $'s with fixed $m_\alpha$'s and 
    (2) optimization of $m_\alpha$'s with fixed $ c^{(\alpha)}_k $'s. 
    The second step is straightforward using \eq{eq:nonoverlapDistribution}.
    For the first step, notice that when $m_\alpha$'s is fixed, the derivatives of $\var{\bar H}$ with respect to $ c^{(\alpha)}_k $'s are linear in $ c^{(\alpha)}_k $'s:  \bea
        \fracpar{\var{\bar H}}{c_k^{(\alpha)}} &=& \nonumber
        \frac{2\sum_{j: \alpha \in \mathcal I_j} c_j^{(\alpha)} \covq{\hat P_k, \hat P_j}{\psi}}{m_\alpha}. 
    \eea 
    To account for the constraints in \eq{eq:cs_constraint}, for each splitting of $c_k$, we fix one of the 
  $\{ c^{(\alpha)}_k \}_{\alpha \in \mathcal I_k}$ 
    as $c^{(*_k)}_k = c_k - \sum_{\alpha \in \mathcal I_k \setminus \{*_k\}} c^{(\alpha)}_k$. 
    The gradients become 
    \bea 
        \fracpar{\var{\bar H}}{c_k^{(\alpha)}} &=& \nonumber
        \frac{2\sum_{j: \alpha \in \mathcal I_j} c_j^{(\alpha)} \covq{\hat P_k, \hat P_j}{\psi}}{m_\alpha} \\ 
        &~& - \frac{2\sum_{j: *_k \in \mathcal I_j} c_j^{(*_k)} \covq{\hat P_k, \hat P_j}{\psi}}{m_{*_k}} \label{eqn:unconstr_grad}
    \eea
    This allows us to find an optimal $ c^{(\alpha)}_k $'s by solving the linear system of equations obtained by equating the gradients to zero. 
    
    If the number of $ c^{(\alpha)}_k $'s overcomes computationally affordable limits, one can always limit the minimization to a selected subset of $ c^{(\alpha)}_k $'s. 
    The criteria for the suitable subset could be the $\hat P_k$ variances, which correlate with magnitudes of their covariances and therefore the importance of their coefficients
    for $\var{\bar H}$. 
    
    \subsection{Optimization by Measurement Allocation}
    \label{sec:MA}

    Another approach to reducing the Hamiltonian estimator variance is to measure each Pauli product as a member of as many compatible measurable fragments as possible. 
    This idea was used in classical shadow tomography methods based on local transformations for measurement of Pauli products.\cite{hadfield2020_locallyBiased,Huang_2021_derand,wu2021overlapped} 
    First, for a particular Pauli product $\hat P_k$, one finds a set of measurement bases $\mathbf B_\alpha$'s where $\hat P_k$ can be measured (see Fig.~\ref{fig:NoO} for an example, by a measurement group this method considers a set of compatible Pauli products). Then, all measurement results for $\hat P_k$ obtained in $\mathbf B_\alpha$'s
    are used to estimate $\bar P_k$: \bea
         \bar P_k &=& \frac{1}{M_k} \sum_{\alpha \in \mathcal I_k }
         \sum_{i=1}^{m_\alpha} P_{k, i}^{(\alpha)},
    \eea where $P_{k, i}^{(\alpha)}$ is the $i$-th measurement result of $\hat P_k$ measured in basis $\mathbf B_\alpha$, 
    and $M_k = \sum_{\alpha \in \mathcal I_k} m_\alpha$ is the total number of times $\hat P_k$ is measured. $\bar P_k$'s are used in the Hamiltonian estimator as $\bar H = \sum_k c_k \bar P_k$.
    The variance of $\bar H$ is \bea
         &&\var{\bar H} = \sum_{jk} c_j c_k \cov{\bar P_j, \bar P_k} \label{eq:overlapVar1}
         \\ &=&
             \sum_{jk}\frac{c_j c_k}{M_j M_k}
             \sum_{\substack{\alpha\in \mathcal I_j,\\ \beta\in\mathcal I_k}}
             \sum_{u=1}^{m_\alpha}
             \sum_{v=1}^{m_\beta}
             \cov{P_{j, u}^{(\alpha)},
             P_{k, v}^{(\beta)}} \label{eq:covHvar}
         \eea 
To proceed further, it is important to distinguish covariances between Pauli products measured within the same fragment and in different fragments. The former correspond to 
$\alpha = \beta$ and $u=v$ in \eq{eq:covHvar} and generally are non-zero, while the latter 
($\alpha \ne \beta$ or $u \ne v$) are zero   
\bea
        \var{\bar H} &=&
             \sum_{jk}\frac{c_j c_k}{M_j M_k}
             \sum_{\substack{\alpha\in \mathcal I_j,\\ \beta\in\mathcal I_k}}
             \sum_{u=1}^{m_\alpha}
             \sum_{v=1}^{m_\beta}
             \delta_{\alpha\beta} \delta_{uv} \covq{\hat P_j, \hat P_k}{\psi}
         \nonumber\\ &=&
             \sum_{jk }\frac{c_j c_k}{M_j M_k}
             \sum_{\alpha \in \mathcal I_j \cap \mathcal I_k } m_\alpha \covq{\hat P_j, \hat P_k}{\psi}.
             \label{eq:overlapVar2}
     \eea
     Note that the key element in deriving this Hamiltonian estimator variance is the consideration that if a Pauli product is measured as a part of a certain group, all members of this group contribute to the average and to the variance. Thus, the variance of each group gives rise to covariances between its members. Since the covariances 
     in different groups are different in magnitude, placing a particular Pauli product in 
     all compatible groups can be sub-optimal for the total variance of the Hamiltonian 
     estimator. 
    
    Dependencies of $M_j$ and $M_k$ on $m_\alpha$'s in $\var{\bar H}$ [\eq{eq:overlapVar2}] 
    make finding the optimal measurement allocation in the analytic form infeasible. 
    To minimize $\var{\bar H}$ with respect to $m_\alpha$'s 
    in \eq{eq:overlapVar2} one can numerically optimize
    $m_\alpha$'s as positive variables with restriction $\sum_\alpha m_\alpha = M$. 
    We will refer to this strategy as the measurement allocation approach. 
    It turns out that such minimization is a version of the coefficient splitting approach with $c^{(\alpha)}_k = c_k m_\alpha/M_k$.
    Indeed, substituting $c^{(\alpha)}_k$'s for $m_\alpha$'s in $\hat A_\alpha$ and using \eq{nonoverlapVar}, we obtain $\var{\bar H}$ as \bea
    \var{\bar H} &=& 
            \sum_\alpha \frac{1}{m_\alpha}\sum_{jk: \alpha \in \mathcal{I}_j \cap \mathcal{I}_k}
                \covq{\frac{m_\alpha}{M_j} c_j \hat P_j, \frac{m_\alpha}{M_k} c_k \hat P_k}{\psi}
            \nonumber 
        \\ &=& 
        \sum_{jk }\frac{c_j c_k}{M_j M_k}
        \sum_{\alpha \in \mathcal I_j \cap \mathcal I_k } m_\alpha \covq{\hat P_j, \hat P_k}{\psi}, 
        \label{eq:overlapVar3} 
    \eea which agrees with \eq{eq:overlapVar2}. 
    Therefore, the measurement allocation approach can be seen as a particular choice of 
   coefficient splitting for reducing $\var{\bar H}$. The main advantage of the 
   measurement allocation approach is a much lower number of optimization variables ($m_\alpha$) compared to that of the coefficient splitting scheme ($c^{(\alpha)}_k$). 
    
    One can formulate approximation for gradients of $\var{\bar H}$ 
    with respect to continuous proxy of $m_\alpha$'s (see Appendix \ref{app:ma_gradient}), which leads to a gradient descent scheme that we will refer to as 
    gradient-based measurement allocation (GMA). 
    Yet, a computationally more efficient, non-gradient iterative 
    scheme was found and detailed below. 

{\it Iterative measurement allocation (IMA): }
    Given an initial guess for $m_\alpha^{(0)}$'s and resulting $M_k^{(0)}$'s, the corresponding coefficient splitting partitioning of the Hamiltonian is
    \bea
        \hat H &=& \sum_\alpha^L \hat A^{(0)}_\alpha, 
    \eea
    where
    \bea
        \hat A^{(0)}_\alpha &=& \sum_{k}
        \frac{m_\alpha^{(0)}}{M_k^{(0)}} c_k \hat P_k,\ \hat P_k \in \mathcal P_\alpha.
    \eea
    Recall that the optimal measurement allocation for any coefficient splitting is given 
    by \eq{eq:nonoverlapDistribution}. Thus, we use this optimal allocation 
    to update $m^{(i)}_\alpha$'s as 
    \bea
        m_\alpha^{(i)} &\rightarrow& m_\alpha^{(i+1)} \propto \sqrt{\varq{\hat{A}^{(i)}_\alpha}{\psi}}, 
    \eea
    which leads to the update in measurable groups
    \bea
        \hat A^{(i)}_\alpha &\rightarrow& \hat A^{(i+1)}_\alpha = \sum_{k}
        \frac{m_\alpha^{(i+1)}}{M_k^{(i+1)}} c_k \hat P_k,\ \hat P_k \in \mathcal P_\alpha
    \eea 
    Since there is no guarantee that each iteration will necessarily lower $\var{\bar H}$ 
    in \eq{eq:overlapVar2}, we repeat these steps multiple times 
    and choose $m_\alpha$'s that result in the lowest estimator variance.
    Empirically, the procedure finds the best measurement allocation in first few cycles.

    \section{Results and Discussions}

    We assess the performance of the proposed approaches 
(IMA, GMA, and ICS) in comparison to prior works (LF, SI, and  classical-shadow-based algorithms) 
    in estimating energy expectation values for ground eigen-states of several molecular electronic Hamiltonians. \cite{app:ham_details}
    To compare different schemes, we normalize the total measurement 
    budget $\sum_\alpha m_\alpha = 1$ and require $m_\alpha$'s to be positive real numbers.
    The estimator variance with $m_\alpha$'s, $\var{\bar H}$, can be scaled by $1/M$ 
    to approximate the estimator variance $\text{Var}'\left(\bar H\right) \approx \var{\bar H}/M$ for $m'_\alpha = \lfloor M m_\alpha\rfloor$ measurements for each group. 
    The overlapping groups ($\mathcal P_\alpha$'s) of the proposed methods are obtained from an extension of the SI technique
    (see Appendix \ref{app:si_extensions}). 
    The initial measurement allocations or 
    coefficient splittings are derived from measurement allocations of the SI technique
    using exact or CISD wavefunctions. 
    
To illustrate the relative performance of our methods, Table~\ref{tab:result_fci} presents the Hamiltonian estimator variances based on covariances calculated with the exact wavefunction. Lower variances in SI compared to those in LF 
are consistent with earlier findings \cite{Crawford2021efficientquantum}. 
All proposed methods result in lower variances than those in SI. 
As the most flexible approach, the coefficient splitting method ICS achieves the lowest variances. 
GMA has a slight edge over IMA in estimator variances, but due to the computational cost of GMA, we will only consider IMA from here on. 

Table~\ref{tab:variable_comparison} shows the number of optimization variables in the measurement allocation and coefficient splitting techniques. For the measurement allocation approaches (IMA and GMA) the number of variables is equal to the number of measurable groups. For the qubit-wise (full) commutativity, the number of such groups scales as $\sim N_P/3$ ($\sim N_q^3$) since on average each group contains three ($N_q$) Pauli products.
For relatively small molecules in our set (i.e. only few atoms), $N_P$ scales as $N_q^4$. In the coefficient splitting approach, the number of variables is a product of $N_P$ and an average number of measurable groups that are compatible with an average Pauli product. 
For our model systems, it was found empirically that the latter number grows as $\sim N_q^3$ for the qubit-wise commutativity, whereas for the full commutativity the number is within 
a range of $[0.4,2.3]$ and thus can be considered relatively constant. 
These considerations clarify why the measurement allocation techniques can be employed for both commutativities,
but the coefficient splitting without extra constraints can be afforded only for the FC grouping. 
    
    \begin{table}[!ht]
        \caption{Variances of the Hamiltonian estimators in different methods calculated with the exact wavefunction. Covariances calculated with the exact wavefunction were used for finding optimal parameters in all methods.  
        }
        \centering
        {\begin{tabularx}{\columnwidth}{@{\extracolsep{\fill}} l c c c c c}
            \toprule
             Systems  & LF & SI  & IMA & GMA & ICS\\
            \midrule
            \multicolumn{6}{c}{Qubit-wise commutativity} \\
             H$_2$   & 0.136 & 0.136 & 0.136 & 0.136 & 0.136 \\
             LiH     & 5.84 & 2.09 & 1.73 & 1.52  & 0.976\\
              BeH$_2$ & 14.3 & 6.34 & 5.60 & 5.26 & 4.29 \\
              H$_2$O  & 116 & 48.6 & 27.9 & 18.8  & 13.5\\
              NH$_3$  & 352 & 97.0 & 83.3 & 62.1  & 44.8\\
            \midrule
            \multicolumn{6}{c}{Full commutativity} \\
             H$_2$   & 0.136 & 0.136 & 0.136 & 0.136 & 0.136 \\
            LiH     & 1.43 & 0.882 & 0.647 & 0.517 & 0.232 \\
             BeH$_2$ & 5.18 & 1.11 & 1.02 & 0.974 & 0.459\\
             H$_2$O  & 43.4 & 7.59 & 5.88 & 4.27  & 1.50 \\
             NH$_3$  & 78.7 & 18.8 & 13.6 & 9.35  & 3.32 \\
            \bottomrule
            \end{tabularx}
        }
        \label{tab:result_fci}
    \end{table}
    
            \begin{table}[!ht]
        \caption{The number of optimization variables in the measurement allocation (MA) 
        and coefficient splitting (CS) methods for the full and qubit-wise commutativities (FC and QWC) and  different molecular electronic Hamiltonians ($N_q$ is the number of qubits, and $N_P$ is the number of Pauli products).  
        }
        \centering
        {\begin{tabularx}{\columnwidth}{@{\extracolsep{\fill}}l c c c c c c}
            \toprule
              \multirow{2}{*}{Systems} & \multirow{2}{*}{$N_q$} & \multirow{2}{*}{$N_P$}  &\multicolumn{2}{c}{QWC}   &       \multicolumn{2}{c}{FC} \\
              \cline{4-5}\cline{6-7}
                \multicolumn{3}{c}{ } & MA & CS & MA & CS\\
            \midrule
             H$_2$   & 4 & 15 & 3 & 4 & 2 & 6\\
             LiH     & 12 & 631 & 155 & 3722 & 42 & 1466\\
             BeH$_2$ & 14 & 666 & 183 & 5946 & 36 & 1203\\
             H$_2$O  & 14 & 1086 & 334 & 11192 & 50 & 1823\\
             NH$_3$  & 16 & 3609 & 1359 & 61137 & 122 & 6138\\
            \bottomrule
            \end{tabularx}
        }
        \label{tab:variable_comparison}
    \end{table}
        

    To compare the proposed methods to the classcial shadow tomography techniques (Derand\cite{Huang_2021_derand} and OGM\cite{wu2021overlapped}), 
    we consider QWC grouping methods that do not require non-local (entangling) transformations (Table~\ref{tab:result_qwc}). Unlike the original OGM treatment,
    we avoid deleting measurement bases to compare all methods on an equal footing. 
    Comparison between the non-overlapping techniques (LF and SI) and classical shadow techniques reveals that only employing the greedy approach to QWC grouping in SI is already enough to surpass the classical shadow tomography techniques. 
    In accord with results of Table~\ref{tab:result_fci}, both IMA and ICS outperform SI 
    even when approximate covariances are used. 


    Lastly, we present improvements on the variances that can be achieved by adopting non-local unitary transformations and FC grouping (Table~\ref{tab:result_fc}).
    For all considered systems, IMA and ICS are superior to SI. This shows that approximate covariances based on the 
    CISD wavefunction can perform similarly to the exact ones. 


    \begin{table*}[!htbp]
        \setlength\tabcolsep{0pt}
        \caption{Variances of Hamiltonian estimators with qubit wise commuting fragments: largest first (LF), overlapped grouping measurement (OGM), 
        derandomization (Derand), sorted insertion (SI), iterative measurement allocation (IMA), and iterative coefficient splitting (ICS).
         The LF, SI, IMA, and ICS algorithms utilize CISD wavefunctions for evaluating covariances,
        but the final variances are computed using exact wavefunctions.  
        }
        \centering
        {\begin{tabular*}{\textwidth}{@{\extracolsep{\fill}}l c c c c c c }
            \toprule
            Systems & LF & OGM & Derand & SI  & IMA & ICS\\
            \midrule
            H$_2$   & 0.136 & 0.173 & 0.144 & 0.136 & 0.136 & 0.136  \\
            LiH     & 5.84 & 3.50 & 3.74 & 2.09 & 1.73 & 0.978  \\
            BeH$_2$ & 14.3 & 18.3 & 12.5 & 6.34 & 5.60 & 4.40 \\
            H$_2$O  & 166 & 148  & 114  & 48.6 & 27.9 & 13.8  \\
            NH$_3$  & 500 & 305  & 251  & 97.0  & 83.4 & 45.5 \\
            \bottomrule
            \end{tabular*}
        }
        \label{tab:result_qwc}
    \end{table*}

    \begin{table*}[!htbp]
        \setlength\tabcolsep{0pt}
        \caption{Variances of Hamiltonian estimators with fully commuting fragments:  largest first (LF), sorted insertion (SI), iterative measurement allocation (IMA), and iterative coefficient splitting (ICS). 
        All algorithms utilize CISD wavefunctions for evaluating covariances,
        but the final variances are computed using exact wavefunctions. 
        }
        \centering
        {\begin{tabular*}{\textwidth}{@{\extracolsep{\fill}}l c c c c}
            \toprule
            Systems & LF & SI & IMA & ICS\\
            \midrule
            H$_2$   & 0.136 & 0.136 & 0.136 & 0.136 \\
            LiH     & 1.43 & 0.882 & 0.647 & 0.232 \\
            BeH$_2$ & 5.19 & 1.11 & 1.02 & 0.495 \\
            H$_2$O  & 43.4 & 7.59 & 5.89 & 1.68 \\
            NH$_3$  & 78.8 & 18.8 & 13.7 & 3.42 \\
            \bottomrule
            \end{tabular*}
        }
        \label{tab:result_fc}
    \end{table*}

    \section{Conclusions}
We assessed multiple ideas for reduction of the number of measurements required to accurately obtain the expectation value of any operator that can be written as a sum of Pauli products. Since these ideas can be used separately or combined, our main goal was to understand the impact on the number of measurements and incurred computational cost of each idea. 
Exploring the idea of Pauli products' compatibility led to the realization that the coefficient splitting framework is the most general implementation of this idea for the grouping methods.   
    
We found that although classical shadow techniques have shown performance superior to that of the non-overlapping measurement scheme based on graph-coloring algorithms, by employing a greedy heuristic the non-overlapping scheme can already outperform the classical shadow techniques. Due to the dependence of the total number of measurements on the sum of square roots of variances for measurable fragments, the greedy approach to grouping performs the best by creating distributions of fragments that are highly non-uniform in variance. The SI technique based on greedy grouping and using only qubit-wise commutativity surpasses shadow tomography based techniques (Derand and OGM) by a factor of 2-3 for larger systems. Thus, for future developments, the shadow tomography approaches need to be compared with greedy grouping based algorithms rather than with grouping approaches that try to minimize the overall number of measurable groups (e.g. LF).     
    
Unlike previous classical shadow techniques that focus on qubit-wise commuting groups, we also considered measuring techniques involving non-local (entangling) transformations that allow one to measure groups of fully commuting Pauli products. An efficient implementation of these non-local transformations using Clifford gates was proposed by Gottesman\cite{Gottesman:CG} and would introduce only $O(N_q^2/\log N_q)$ CNOT gates. The schemes based on 
fully commuting groups outperform their qubit-wise commuting counterparts up to a factor of seven in variances of the expectation value estimators. 
    
Taking advantage of compatibility of some Pauli products with members of multiple measurable groups (i.e. overlapping groups idea) can be generally presented as augmenting the measurable groups with all Pauli products compatible with initial members of these groups. Then the coefficients of Pauli products entering multiple groups are optimized to lower the estimator variance, with the constraint that the sum over coefficients in different groups for each Pauli product is equal to the coefficient of the Pauli product in the Hamiltonian. This coefficient splitting approach incorporates as a special case 
a heuristic technique of optimizing measurement allocation for overlapping measurable groups.

Even though the coefficient splitting variance minimization provides the lowest variances among all studied approaches, it requires optimizing a large number of variables: $\sim N_q^4$ ($\sim N_q^7$) for full (qubit-wise) commutativity. Due to certain restrictions, the measurement allocation approach is much more economical in the number of optimization variables: $\sim N_q^3$ ($\sim N_q^4$) for full (qubit-wise) commutativity. Another 
contributor of the computational cost of these techniques is calculation of the 
variance gradients. To reduce the computational cost of this part we proposed iterative schemes, the ICS method converges to true extrema, while the IMA scheme deviates from 
extrema. IMA and ICS provide up to forty and eighty percent reduction in the number of measurements required compared to corresponding best non-overlapping techniques. 

Both IMA and ICS use approximate covariances between Pauli products 
to lower the estimator variance. Use of CISD wavefunction for calculating these 
covariances shown improvements comparable to those obtained using the exact covariances. 
Additionally, in IMA and ICS, one can improve covariances obtained from approximate wavefunctions 
using accumulated measurement results.

    \section*{Data Availability}
    The data that support the findings of this study are 
    available from the corresponding author upon request.
    
    \section*{Code Availability}
    Some part of the code that supports the findings of this study is available in the OpenFermion\cite{OpenFermion} and PySCF\cite{PySCF} libraries. The rest of the code is available 
    from the corresponding author upon request.
    
    \section*{Acknowledgements}
    T.Y. is grateful to Hsin-Yuan Huang and Bujiao Wu for providing details of their respective algorithms in Refs. \citenum{Huang_2021_derand} and \citenum{wu2021overlapped}. The authors thank Seonghoon Choi for useful discussion. A.F.I. acknowledges financial support from the Google Quantum Research Program, Early Researcher Award, and Zapata Computing Inc. This research was enabled in part by support provided by Compute Ontario and Compute Canada.
    
    \section*{Author Contributions}
    T.-C.Y. and A.F.I conceptualized the project and wrote most of the paper. A.G. developed IMA and collected all the data, except for calculations of LF (in Table.~\ref{tab:result_fci}), SI (in Table.~\ref{tab:result_fci}), GMA, and OGM that T.-C.Y. performed.  T.-C.Y., A.G., and A.F.I participated in discussions that developed the theory as well as the GMA and ICS methods. 
    T.-C.Y. and A.G. share co-first authorship. 
        
    \section*{Competing Interests}
    The authors declare that there are no competing interests.

    \appendix
    
    \section{Gradients of Measurement Allocation Optimization}
    \label{app:ma_gradient}
    
    The variance of the estimator for $\braket{\hat H}$ as a function of $m_\alpha$'s and $M_k$'s is 
    \bea
        \var{\bar H} &=& \sum_{ij}\frac{c_i c_j}{M_i M_j}
            \sum_{\alpha \in \mathcal I_i \cap \mathcal I_j} m_\alpha \covq{\hat P_i, \hat P_j}{\psi}
            \eea
            \bea
     &=&
            \sum_{k} \frac{c_k^2}{M_k} \text{Var}\left(\hat P_k \right)
        \nonumber\\&+& \sum_{i > j} \sum_{\alpha \in \mathcal I_i \cap \mathcal I_j}
            \frac{2m_\alpha}{M_i M_j} c_i c_j
            \covq{\hat P_i, \hat P_j}{\psi}.
    \eea
    Note that $M_i$'s are dependent on $m_\alpha$'s.
    We minimize $\var{\bar H}$ using $m_\alpha$'s as variables under the constraints $\sum_\alpha m_\alpha = 1$ and $m_\alpha > 0$. 
    The variance derivatives with respect to $m_\alpha$'s are \bea
        \fracpar{\var{\bar H}}{m_\alpha} &=&
            \sum_{k: \alpha \in \mathcal I_k} -\frac{c_k^2}{M_k^2} \text{Var}\left(\hat P_k \right)\notag
        \\ &+&
            \sum_{\substack{i>j:\\ \alpha \in \mathcal{I}_i \cap \mathcal{I}_j}}\left(
                \frac{1}{M_i M_j}
                - \frac{m_\alpha}{\left(M_i\right)^2 M_j}
                - \frac{m_\alpha}{M_i \left(M_j\right)^2 }
            \right)
            \nonumber
            \\ &\times&
            2c_i c_j
            \covq{\hat P_i, \hat P_j}{\psi} \notag
        \\ &+&
            \sum_{i > j} \sum_{\substack{\beta \neq \alpha:\\ \beta \in \mathcal{I}_i \cap \mathcal{I}_j}}
            \left( \begin{cases}
                - \frac{m_\beta}{\left(M_i\right)^2 M_j}
                & \text{if } \alpha \in \mathcal I_i \\
                - \frac{m_\beta}{M_i \left(M_j\right)^2}
                & \text{if } \alpha \in \mathcal I_j \\
                \text{1st} + \text{2nd} & \text{if } \alpha \in \mathcal I_i \cap \mathcal I_j
            \end{cases} \right)
            \nonumber \\ &\times&
            2c_i c_j
            \covq{\hat P_i, \hat P_j}{\psi}. \label{eq:Vm}
    \eea
    To avoid constrained optimization with $m_\alpha$'s, we use auxiliary variables
    $p_\alpha$'s that express $m_\alpha$'s as
    \bea
        m_\alpha = \frac{e^{p_\alpha}}{\sum_\beta e^{p_\beta}}
    \eea
    to introduce the $\sum_\alpha m_\alpha = 1$ and $m_\alpha > 0$ conditions. 
    This is known as the softmax function often used in machine learning techniques.
    Derivatives $\partial \var{\bar H}/ \partial p_\beta$ only require additional terms
    \bea
        \fracpar{m_\alpha}{p_\beta} = \begin{cases}
             m_\beta\sparen{1 - m_\beta} & \alpha = \beta \\
             -m_\alpha m_\beta & \alpha \neq \beta
        \end{cases}
    \eea
    for completing a chain-rule expression with \eq{eq:Vm}.  
    
    \section{Extensions of Sorted Insertion}
    \label{app:si_extensions}

    Sorted Insertion (SI)\cite{Crawford2021efficientquantum} is one of the most efficient measurement schemes that utilizes non-overlapping Pauli groups.
    Here, we briefly review the original implementation and introduce modifications to find overlapping groups for the coefficient splitting and measurement allocation approaches. 
    
    SI partitions all the Pauli products $\mathcal{H} = \{ \hat P_k\}$ in $\hat H$ into a set of non-overlapping groups $\mathbf{G} = \{\mathcal P_\alpha\}$ such that \bea
        \hat H &=& \sum_k c_k \hat P_k = \sum_\alpha \hat A_\alpha \\
        \hat A_\alpha &=& \sum_{\hat P_k^{(\alpha)} \in \mathcal P_\alpha} c_k \hat P_k^{(\alpha)}.
    \eea
    SI initiates $\mathbf{G} = \emptyset$, $\alpha=1$ and finds the partitioning through the following steps:
    \begin{enumerate}
        \item Sort Pauli products in $\mathcal H$ in the descending order of the magnitudes of their coefficients.
        \item Examine each product $\hat P \in \mathcal{H}$.
        If $\hat P$ commutes with all products in $\mathcal P_\alpha$, \bea
            \mathcal P_\alpha &\rightarrow& \mathcal P_\alpha \cup \{\hat P\}  \\
            \mathcal H &\rightarrow& \mathcal H \setminus \{\hat P\}.
        \eea
        \item Add $\mathcal P_\alpha$ to $\mathbf G$. Set $\alpha \rightarrow \alpha+1$ and repeat from step 2 until $\mathcal H$ is empty.
    \end{enumerate}

    In order to obtain overlapping groups, we maintain set $\mathcal P_*$ to track Pauli products that are already part of some fragments, and examine whether they are compatible with the group $\mathcal P_\alpha$ built in step 2. 
    Note that the order in which the Pauli products are added to the groups matters, since the additional Pauli products that are compatible with the SI groups may not be compatible between themselves. 
    We initiate $\mathcal P_* = \emptyset$ and add an extra procedure between step 2 and 3: \begin{itemize}
        \item
        For $\hat P \in \mathcal P_*$ in the order they were added to $\mathcal P_*$,
        add $\hat P$ to $\mathcal P_\alpha$ if $\hat P$ is compatible with all members of $\mathcal P_\alpha$.
        Then, $\forall \hat P \in \mathcal P_\alpha\setminus \mathcal P_*$, in the order added to $\mathcal P_\alpha$, set $\mathcal P_* \rightarrow \mathcal P_* \cup \{\hat P\}$
    \end{itemize}


\begin{thebibliography}{30}
\bibitem [{app()}]{app:ham_details}%
  \BibitemOpen
  \href@noop {} {}\bibinfo {note} {The qubit Hamiltonians were generated using
  the STO-3G basis and the BK transformation. The nuclear geometries for the
  Hamiltonians are R$(\rm H-H) = 1 \AA$ ($\rm H_2$), R$(\rm Li-H) = 1 \AA$
  ($\rm LiH$), R$(\rm Be-H) = 1 \AA$ with collinear atomic arrangement ($\rm
  BeH_2$), R$(\rm O-H) = 1 \AA$ with $\angle HOH = 107.6^\circ$ ($\rm H_2O$),
  and R$(\rm N-H) = 1 \AA$ with $\angle HNH = 107^\circ$ ($\rm
  NH_3$).}\BibitemShut {Stop}%
\end{thebibliography}
%

\end{document}